\documentclass[twocolumn,showpacs,preprintnumbers,amsmath,amssymb]{revtex4}

\usepackage{graphicx}
\usepackage{dcolumn}
\usepackage{bm}

\newcommand{\beq}{\begin{equation}}
\newcommand{\bqq}{\begin{eqnarray}}
\newcommand{\eqq}{\end{eqnarray}}

\begin{document}

\title{Mechanism of proton-$^3{\rm He} $ elastic backward
scattering at intermediate energy}

\author{A.P.~Kobushkin}
 \altaffiliation[Also at ]{Bogolyubov Institute for Theoretical Physics,
Metrologicheskaya str. 14B, 03143 Kiev, Ukraine}
\email{akob@rcnp.osaka-u.ac.jp}
\author{E.A.~Strokovsky}%
 \altaffiliation[Also at ]{LPP, Joint Institute for Nuclear Research,
141980, Dubna, Moscow region, Russia}
\email{strok@rcnp.osaka-u.ac.jp}
\author{K.~Hatanaka}
\email{hatanaka@rcnp.osaka-u.ac.jp}
\affiliation{Research Center for Nuclear Physics, Osaka University,
10-1 Mihogaoka, Ibaraki, Osaka, 567-0047, Japan}
\author{S.~Ishikawa}
\affiliation{Hosei University, Department of Physics, Fujimi 2-17-1,
Chiyoda, Tokyo, 102-8160, Japan}

\date{\today}
\begin{abstract}
We provide systematic analysis of the
proton-$^3{\rm He}$ elastic scattering at $\theta_{\rm cm}=180^{\circ}$
and at the incident proton energy $T_p\lesssim$~700~MeV. Three mechanisms are
discussed: 2N pair exchange in the triplet and singlet spin states, pion-exchange
and direct mechanism. It is shown that at $T_p\gtrsim~150~\mathrm{MeV}$
three-body structure, including triplet and singlet scattering states of the 2N pair,
becomes of great importance for understanding energy dependence of the reaction 
observables. Predictions for the
differential cross section and the polarization correlation $C_{00nn}$
which can be studied now in experiment are given.
\end{abstract}

\maketitle

\tolerance=10000          

\section{Introduction}
For several decades considerable efforts have been done to
investigate structure of the lightest nuclei (the deuteron,
$^3\rm He$, $^4\rm He$) at
short distances between the constituent nucleons. Significant
progress was achieved both in theory and experiment, first of all
because high quality data on spin-dependent observables were
obtained with both  hadronic \cite{DEBS,Azhgirey} and
electromagnetic probes \cite{em_prob}.
Large part of these investigations consists
of study of elastic backward (in the center of mass system) proton --
nucleus scattering (EBS). This process involves large momentum
transfer  $|t|$ and therefore a belief exists that
EBS can provide an access to the high momentum components of the
wave function of the lightest nuclei.

From those investigations it becames obvious that at present there is no
theoretical model which quantitatively
describes the existing data, even for the simplest reaction,
$pd$ EBS (see \cite{Azhgirey} and Refs. therein). Surprisingly,
the wide gap exists between precise and detailed data base 
and rather imprecise (even qualitatively) theoretical understanding of these reactions.

The elastic backward $p(^3{\rm He},p)^3\rm He$ scattering is studied in much
less detail than the $pd$ EBS. But presently, high intensity beams of
polarized protons in combination with polarized $^3\rm He$
targets \cite{target} give an opportunity to perform detailed studies
of $p^3\rm He$ EBS with spin dependent observables at energy between 200 
to 400~MeV \cite{Proposal}. This, in turn, demands careful
theoretical study of the reaction mechanism, specially at intermediate energy.

The goal of the present study is to develop a theoretical description of
the proton EBS off the lightest nuclei, which provides 
predictions for experimentally measurable observables including spin-dependent
ones. In the present paper the \mbox{$p ^3$He} EBS up to $T_p\sim 0.7$~GeV
is considered only.
In this particular case a request from experiment is to find an adequate
connection of this reaction with the structure of the 3N system and
to get
quantitative estimations for sensitivities of its cross section and
spin-dependent observables to the existing wave functions of $^3\rm He$.

\begin{figure*}
\includegraphics{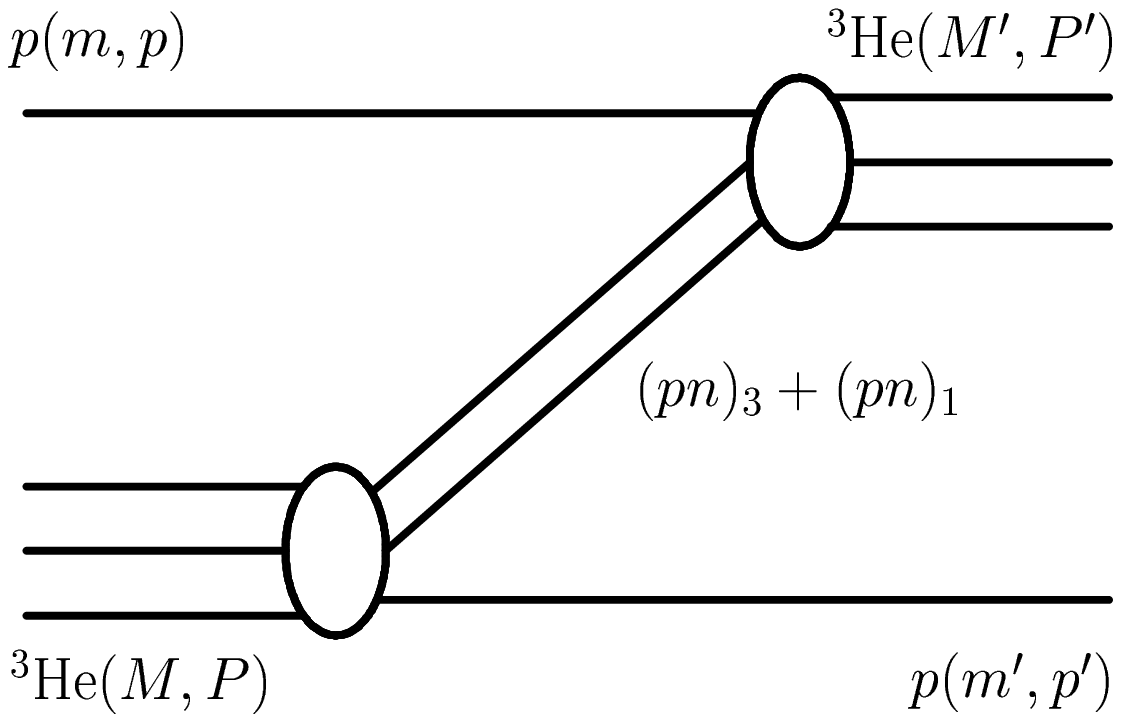}
\includegraphics{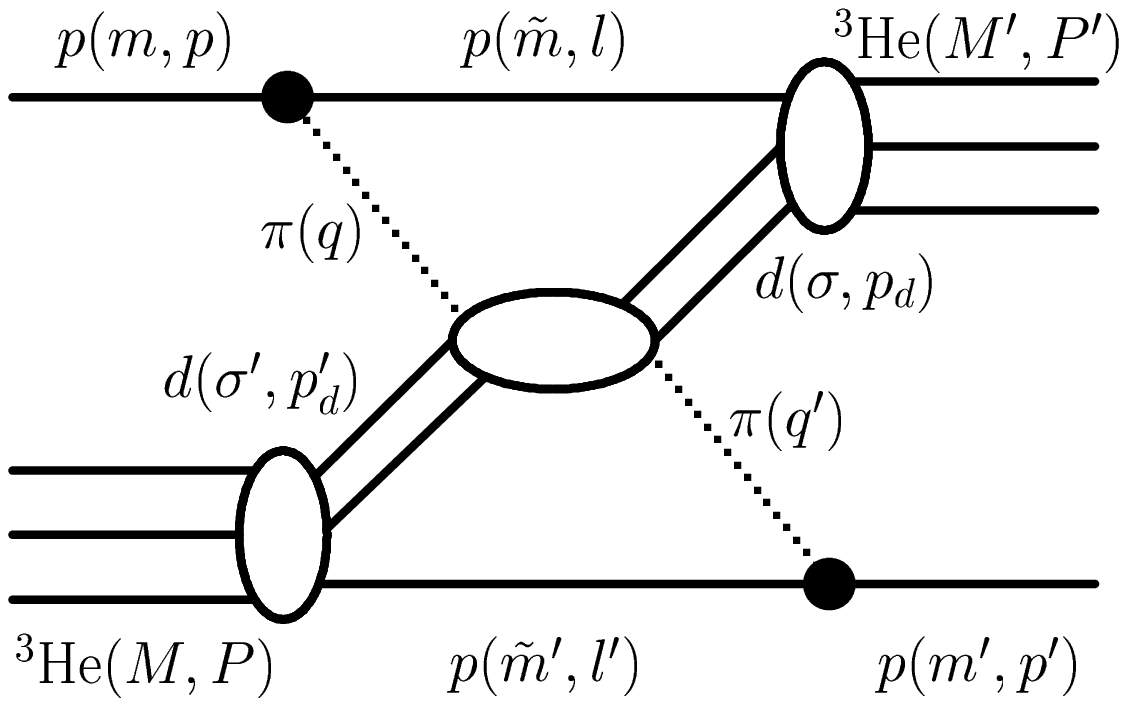}

\includegraphics{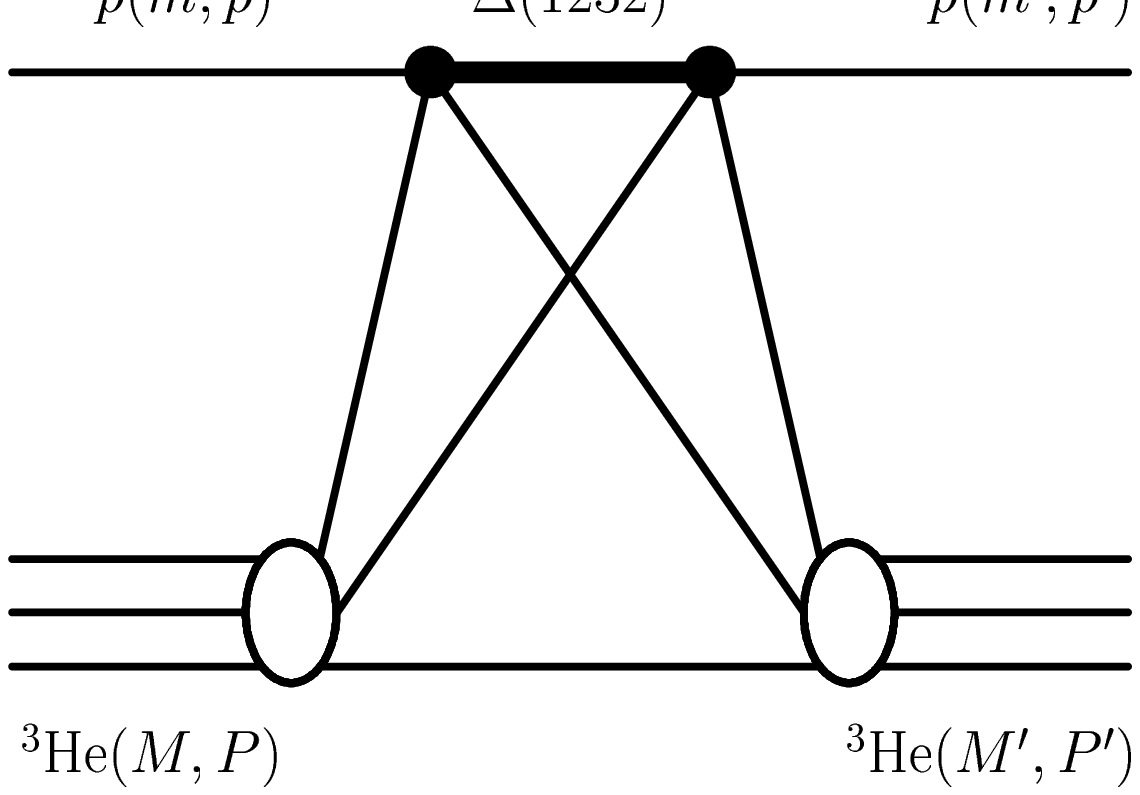}
\includegraphics{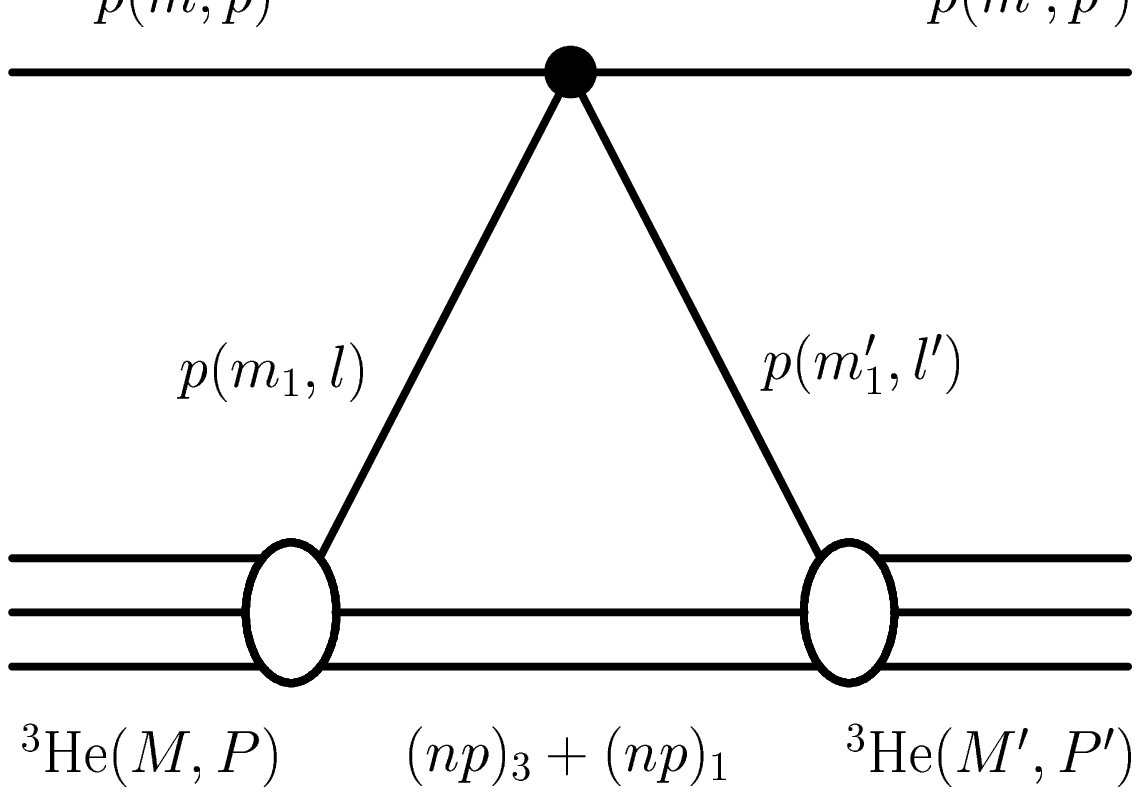}

\vspace{11.cm}
\caption{Mechanisms of the proton-$^3\mathrm{He}$ backward
scattering at intermediate energy: 2N-exchange (upper-left diagram),
PI (upper-right diagram), the $\Delta$-excitation (bottom-left diagram) 
and DIR (bottom-right diagram).}
\label{fig:1}
\end{figure*}

The mechanism of one-deuteron-exchange  shows that such connection
exists, but it fails to reproduce cross section data at $T_p>200$~MeV 
\cite{Lesniaks}. The exchange of the singlet spin  $np$ pair, in addition
to the one-deuteron-exchange, also does not explain existing data at 
intermediate energy \cite{Kob93}. In the present paper we take into account
full three-body structure of $^3\mathrm{He}$ and consider 2N-exchange in 
the singlet and the triplet spin states (the upper-left diagram of 
Figure~\ref{fig:1}). Note that the one-deuteron-exchange gives only a partial 
contribution in the triplet 2N-exchange.

It was stressed by many authors that the so-called pion (PI) mechanism is
responsible for the ``shoulder'' in the measured  energy dependence of the
differential cross section of $p^3\rm He$ EBS
between 200 and 700 MeV \cite{Berthet}-\cite{Uzikov02}. Following the general 
ideas coming from the similar situation in $pd$-EBS \cite{CWilkin,KolybSm}, 
the authors of Refs.\cite{Berthet}-\cite{Uzikov02} estimated the contribution
of PI-mechanism from the triangle diagram with subprocesses 
$pd\to {^3\mathrm{He}}\pi^0$, $pd^\ast\to {^3\mathrm{He}}\pi$. However,
it was already mentioned that in the triangle diagram the initial
and the final nuclei are coupled in different vertices 
and thus this diagram is not T-invariant \cite{Nakamura}. Moreover,
sometimes it is difficult to avoid double-counting with other mechanisms in such approach.

We estimate PI-mechanism from a two-loop diagram where high momentum from the
initial to the final proton is transferred by a virtual pion scattered off 
the intermediate deuteron ( the upper-right diagram of Figure~\ref{fig:1}). 
For the backward $pd$ scattering similar mechanism (elastic 
$\pi N$ scattering in the region of $\Delta$-resonance)  was discussed in 
\cite{Anjos}. We use results of the partial wave analysis 
of the elastic $\pi d$ scattering done by Virginia group \cite{Arndt},
what makes not necessary to take into account resonance propagator explicitly. 
Some authors propose another way to consider pion degrees of freedom which 
also does not contain the abovementioned  difficulties \cite{Kondratyuk81}.   
In our case it corresponds to the bottom-left diagram of
Figure~\ref{fig:1}. Kinematic estimations show that this diagram 
may contribute significantly near $T_p\sim$~0.5~GeV and thus it is not 
essential at RCNP energy. Alternatively the main contribution of the 
upper-right diagram of Figure~\ref{fig:1} appears earlier, around 
$T_p\sim$~0.3~GeV, exactly at the energy of the experiment \cite{Proposal}.   
It corresponds to the pion energy $T_\pi \approx 0.2$~GeV, where the 
differential cross section of the elastic $\pi d $ scattering has a sharp 
resonance-like structure \cite{Garsilazo}.      
 
Among other important mechanisms, the so-called direct mechanism (DIR, the
bottom-right diagram of Figure~\ref{fig:1}) may also play important role in
this reaction \cite{Gurvitz_et_al,Uzikov98}.

The paper is organized as follows. In Sec.~II the general 
formalism of the reaction is given. Parametrization for the $^3\mathrm{He}$ 
wave function is discussed in Sec.~III. In the same section we 
give also results for two-nucleon momentum distributions and densities 
in singlet and triplet spin states coming from  this parametrization. Then, 
in Sec.~IV, we derive reaction amplitudes for three mechanisms, 
2N-exchange, PI and DIR. Results of numerical calculations, comparison 
with experiment and predictions are presented in Sec.~V. 
Conclusions and remarks are given in Sec.~VI. 

\section{\label{sec:General}General formalism}
Parity and time-reversal invariance leaves only 6 independent complex
amplitudes for the elastic scattering of two spin-$\frac12$ particles.
At $\theta_{\rm cm}=180^{\circ}$ three of them vanish and they are
reduced to
\bqq
{\cal M}(\theta_{\rm cm}=180^{\circ})&=&
\left(
\begin{array}{c c c c}
{\cal M}_{++}^{++} & {\cal M}_{+-}^{++} &{\cal M}_{-+}^{++} &{\cal M}_{--}^{++} \\
{\cal M}_{++}^{+-} & {\cal M}_{+-}^{+-} &{\cal M}_{-+}^{+-} &{\cal M}_{--}^{+-} \\
{\cal M}_{++}^{-+} & {\cal M}_{+-}^{-+} &{\cal M}_{-+}^{-+} &{\cal M}_{--}^{-+} \\
{\cal M}_{++}^{--} & {\cal M}_{+-}^{--} &{\cal M}_{-+}^{--} &{\cal M}_{--}^{--}
\end{array}
\right)
\nonumber \\
& = &\left(
\begin{array}{c c c c}
A & 0 & 0 & 0\\
0 & F & G & 0\\
0 & G & F & 0\\
0 & 0 & 0 & A
\end{array}
\right)
\label{1}
\eqq
Here ${\cal M}_{Mm}^{M'm'}$ is amplitude with $M$, $m$ and $M'$, $m'$ magnetic quantum 
numbers of
$^3{\rm He}$ and the proton in initial and final states, respectively.
We assume the  normalization when the differential cross section is given by
\beq\frac{d\sigma}{d\Omega_{\rm cm}} =\frac{\frac14 {\rm Tr}({\cal MM}^\dag)}{64 \pi^2 s} 
=\frac{|A|^2 + |F|^2 + |G|^2}{128 \pi^2 s},\label{3}\end{equation}
where $s$ is the total c.m. energy squared and the factor $\frac14$
comes from the average over the initial spin states.

Corresponding to the three non-vanishing amplitudes there are
$2 \times 3 -1 =5$ independent observables: the differential cross section
(\ref{3}) and 4 spin-dependent observables. Among the last ones we
will consider in this paper only polarization correlation
\beq
C_{00nn}=\frac{
{\rm Tr}({\cal M}\sigma_y({\rm He}) {\cal M}^\dag\sigma_y(p))
}
{
{\rm Tr}({\cal MM}^{\dag})
}=
\frac{2{\rm Re}( A G^{\ast})}
{|A|^2 + |F|^2 + |G|^2}.
\label{4}
\end{equation}
\section{$^3{\rm He}$ wave function \label{WFunc}}
In forthcoming calculations we use a new parametri\-zation of full an\-ti\-symmetric trinucleon 
wave func\-tion (TWF) \cite{Baru} for Paris \cite{Paris} and CD-Bonn 
\cite{CD-Bonn} potentials. It is restricted by the five partial wave components
\beq
\nu=\left\{
^1s_0S,\ ^3s_1S,\ ^3s_1D,\ ^3d_1S,\ ^3s_1D
\right\},
\label{III.1}
\end{equation}
where $^1s_0$ etc. correspond to a pair of nucleons and $S$ and $D$ denotes relative
angular momentum between the pair and the spectator nucleon.  The components for higher 
angular momenta were shown to be unsignificant. The (1,2) pair is chosen as
an active pair and the TWF components are defined as follows
\bqq
&&\left<r \rho \nu|\Psi\right>=
\left<r \rho \nu|\psi[(12)3]\right> 
\nonumber \\
&& + \sum_{\nu_{23}}\left<r \rho \nu|\psi[(23)1]\right> +\sum_{\nu_{31}}\left<r \rho \nu|\psi[(31)2]\right>,
\label{III.2}
\eqq
where
\beq
\left<r \rho \nu|\Psi\right>=\sum \left<...\right> Y_{ll_3}(\hat r) Y_{LL_3}(\hat \rho)\Psi_\nu(r,\rho)
\label{III.3}
\end{equation}
and  $\left<...\right>$ denotes all necessary spin and isospin Clebsh-Gordan coefficients;
$\vec r=\vec r_{12}$ is the relative coordinate in the pair and $\vec \rho=\vec \rho_3$ is the
relative coordinate between the pair center of mass and the third nucleon. In the momentum 
space $\vec p$ and $\vec q$ are used for the relative momentum in the pair and the
relative momentum between the pair and the third nucleon, respectively.

The normalization is given by
\beq
\int_0^\infty dr d\rho \sum_\nu \left|\Psi_\nu(r,\rho)\right|^2=1.
\label{III.norma}
\end{equation}
Important quantities in our calculations are momentum distributions of a virtual
$(np)$ pair in triplet ($j=1$) and singlet ($j=0$) states
\bqq
n_{3}^S(q)&=&3\int_0^\infty dp p^2 \left[
\left|\Psi_{^3s_1S}(p,q) \right|^2+\left|\Psi_{^3d_1S}(p,q) \right|^2 
\right],
\nonumber \\
n_{3}^D(q)&=&3\int_0^\infty dp p^2 \left[
\left|\Psi_{^3s_1D}(p,q) \right|^2 +
\left|\Psi_{^3d_1D}(p,q) \right|^2 
\right],
\nonumber \\
n_{3}^\mathrm{int}(q)&=&3\int_0^\infty dp p^2 \left[
\Psi_{^3s_1S}(p,q)\Psi_{^3s_1D}(p,q)
\right.\nonumber \\
&&
\left. \ \ \ \ \ \ \ \ \ \ \ \ \
+ \Psi_{^3d_1S}(p,q) \Psi_{^3d_1D}(p,q)
\right],
\nonumber \\
n_{1}(q)&=&3\int_0^\infty dp p^2 
\left|\Psi_{^1s_0S}(p,q) \right|^2,
\label{D_starMD}
\eqq
where the factor 3 is a combinatorial factor and the bottom suffix is $2j+1$.

For PI-mechanism we also need the deuteron momentum distribution in $^3\mathrm{He}$.
It is given by the $S$ and $D$ waves for relative motion in the $d+p$-component of the 
$^3\mathrm{He}$ wave function
\bqq
u(q)&=&\int_0^\infty dp p^2 \left[u_d(p)\Psi_{^3s_1S}(p,q) +
\right.
\nonumber \\
&& \left. \ \ \ \ \ \ \ \ \ \ \ \ + w_d(p)\Psi_{^3d_1S}(p,q)
\right],
\nonumber \\
w(q)&=&\int_0^\infty dp p^2 \left[u_d(p)\Psi_{^3s_1D}(p,q) +\right.
\nonumber \\
&&  \left.\ \ \ \ \ \ \ \ \ \ \ \ +
 w_d(p)\Psi_{^3d_1D}(p,q)
\right],
\label{III.4}
\eqq
where $u_d(p)$ and $w_d(p)$ are the deuteron $S$ and $D$ wave functions, respectively. Now the deuteron
momentum distribution reads
\beq
n_d(q)=n_d^S(q)+n_d^D(q)=3\left[u^2(q) + w^2(q)\right],
\label{III.DMD}
\end{equation} 
where $n_d^S(q)=3u^2(q)$ and $n_d^D(q)=3w^2(q)$ have meaning of the deuteron momentum distributions in
the $S$ and $D$ waves. Partially they contribute to $n_{3}^S(q)$ and $n_{3}^D(q)$, respectively.
The quantity $n_d^\mathrm{int}(q)=3u(q)w(q)$ corresponds to the deuteron contribution in the ``interference'' momentum distribution $n_{3}^\mathrm{int}(q)$.

According to the normalization (\ref{III.norma}) the quantity
\bqq
N_d&=&\int_0^\infty dq q^2\left(n_d^S(q)+n_d^D(q)\right)\nonumber \\
&=&3\int_0^\infty dq q^2
 \left[u^2(q) + w^2(q)\right]
\label{III.5}
\eqq 
has meaning of the effective number of the deuterons in $^3\mathrm{He}$.
We obtain $N_d=1.39$ for Paris potential and $N_d=1.36$ for CD-Bonn potential. This agrees well
with $N_d=1.38$ obtained earlier for Urbana and Argonne potentials \cite{Schiavilla}.

\begin{figure*}
\includegraphics{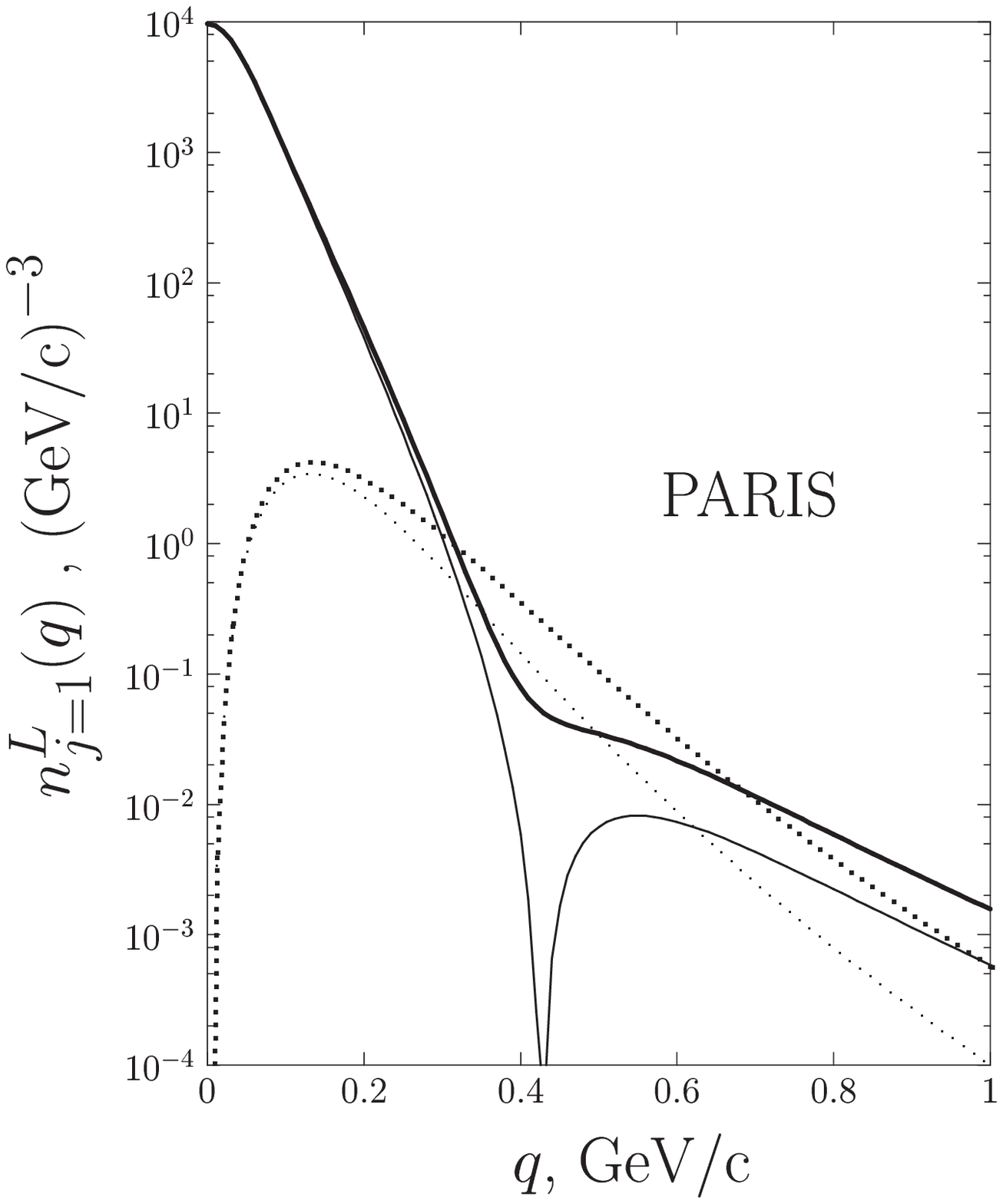}
\includegraphics{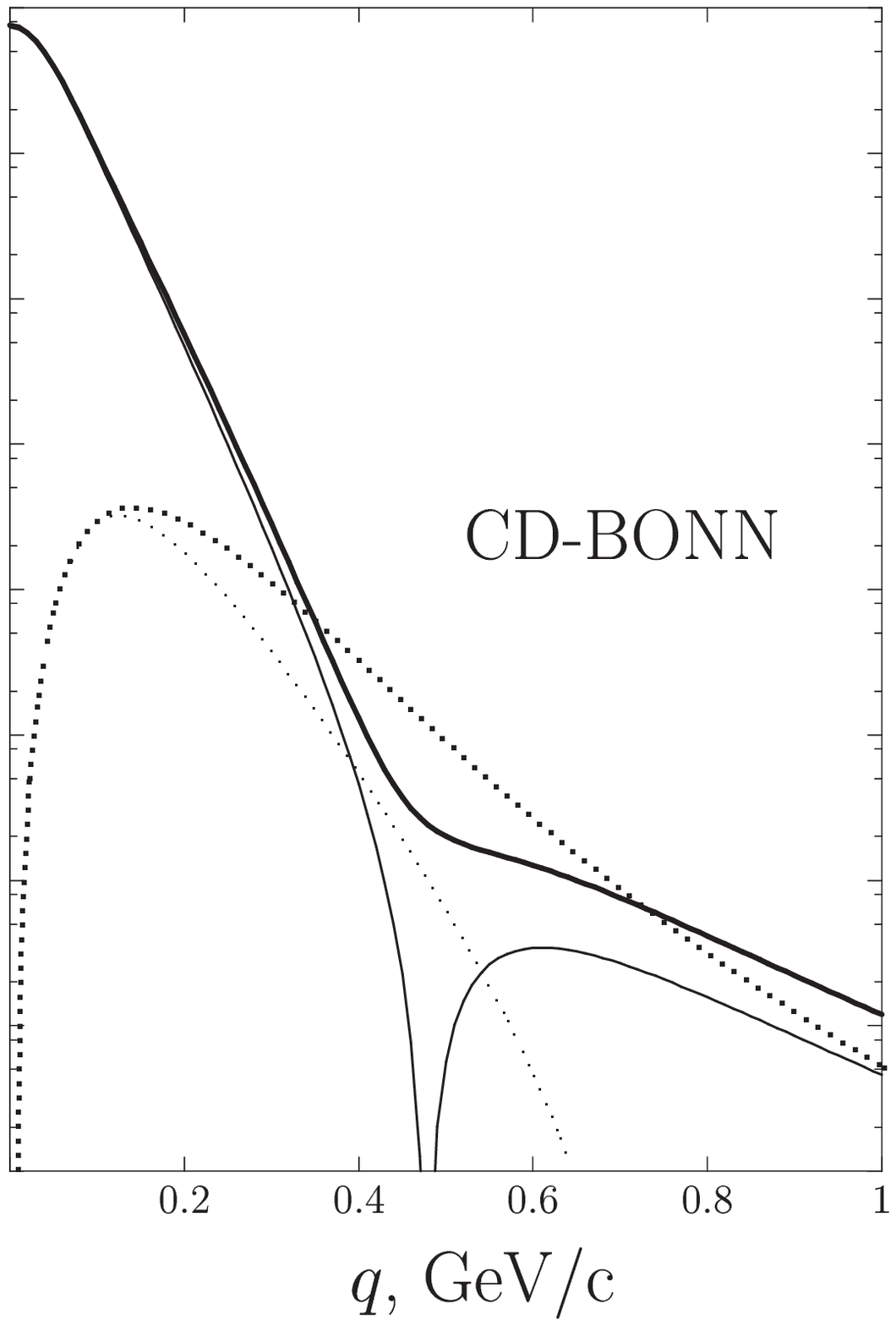}

\vspace{10.cm}
\caption{Momentum distributions of the triplet pair (the full and dotted bold lines are for $S$ 
and $D$ wave, respectively)
and of the the deuteron (the full and dotted thin lines are for $S$ 
and $D$ wave, respectively).}
\label{fig:2:S+D}
\end{figure*}

\begin{figure*}
\includegraphics{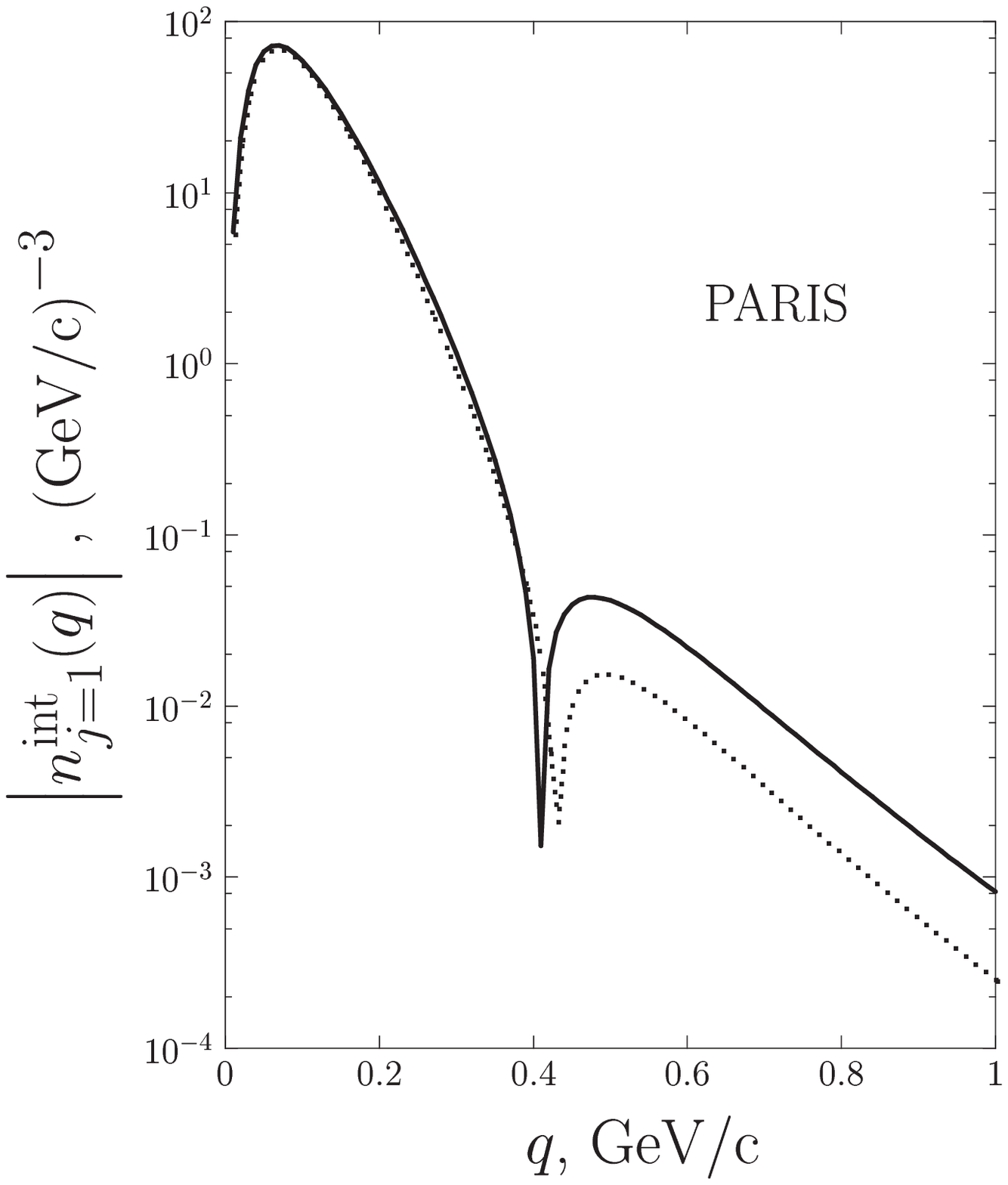}
\includegraphics{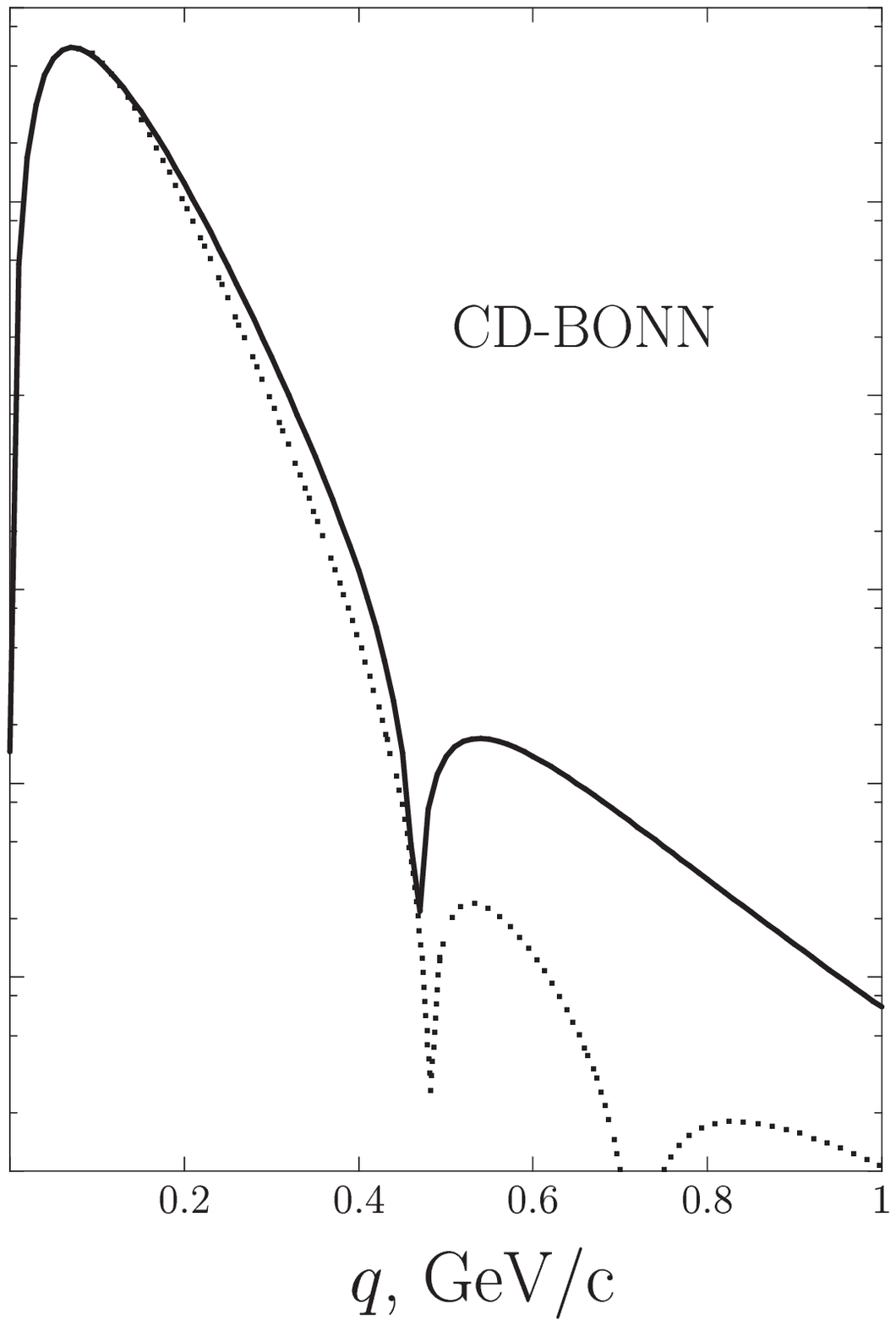}

\vspace{10.cm}
\caption{Interference momentum distribution: the full line is for the triplet pair
$n_{3}^\mathrm{int}(q)$ and the dotted line is for $u(q)w(q)$.}
\label{fig:3:int}
\end{figure*}

\begin{figure*}
\includegraphics{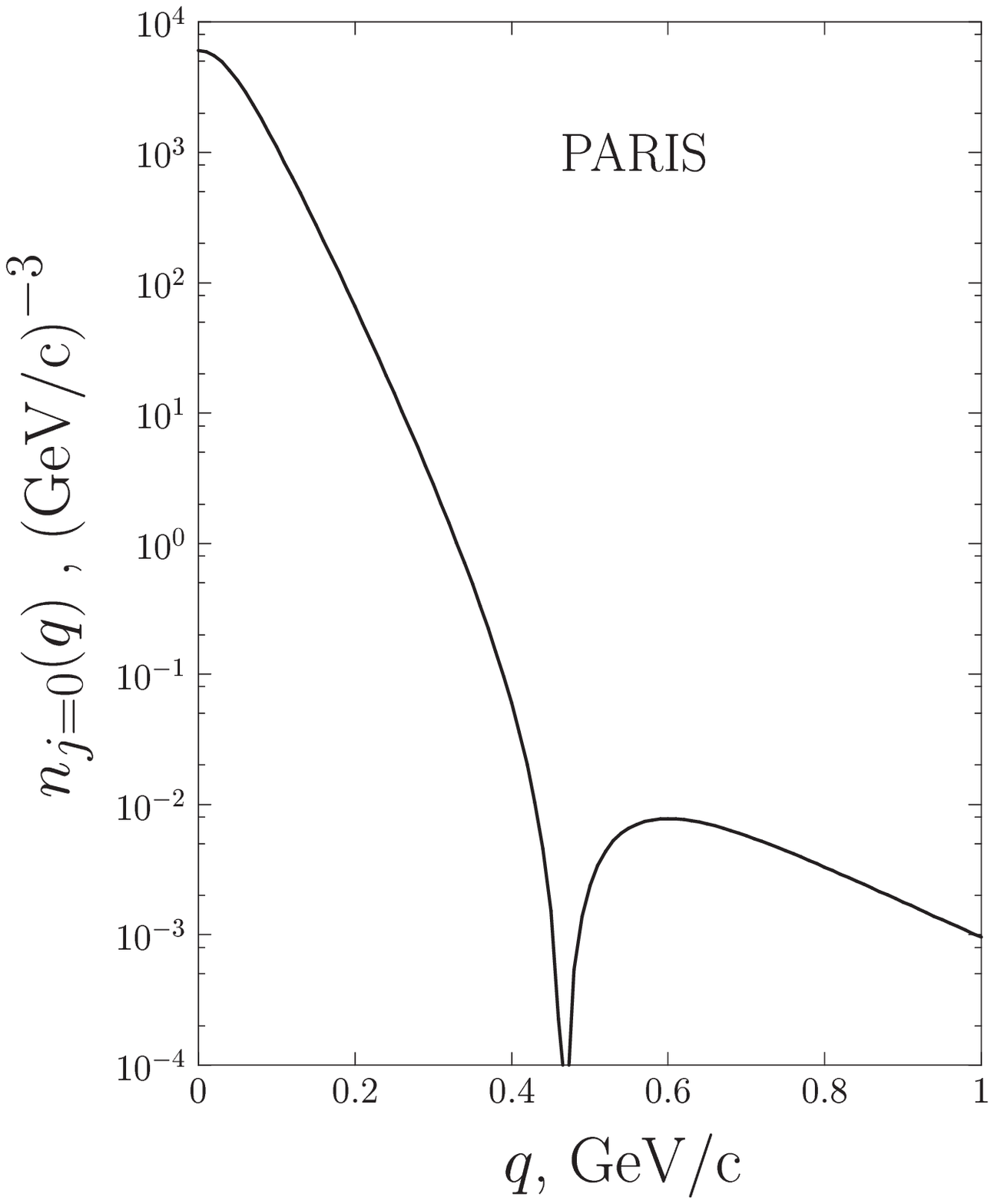}
\includegraphics{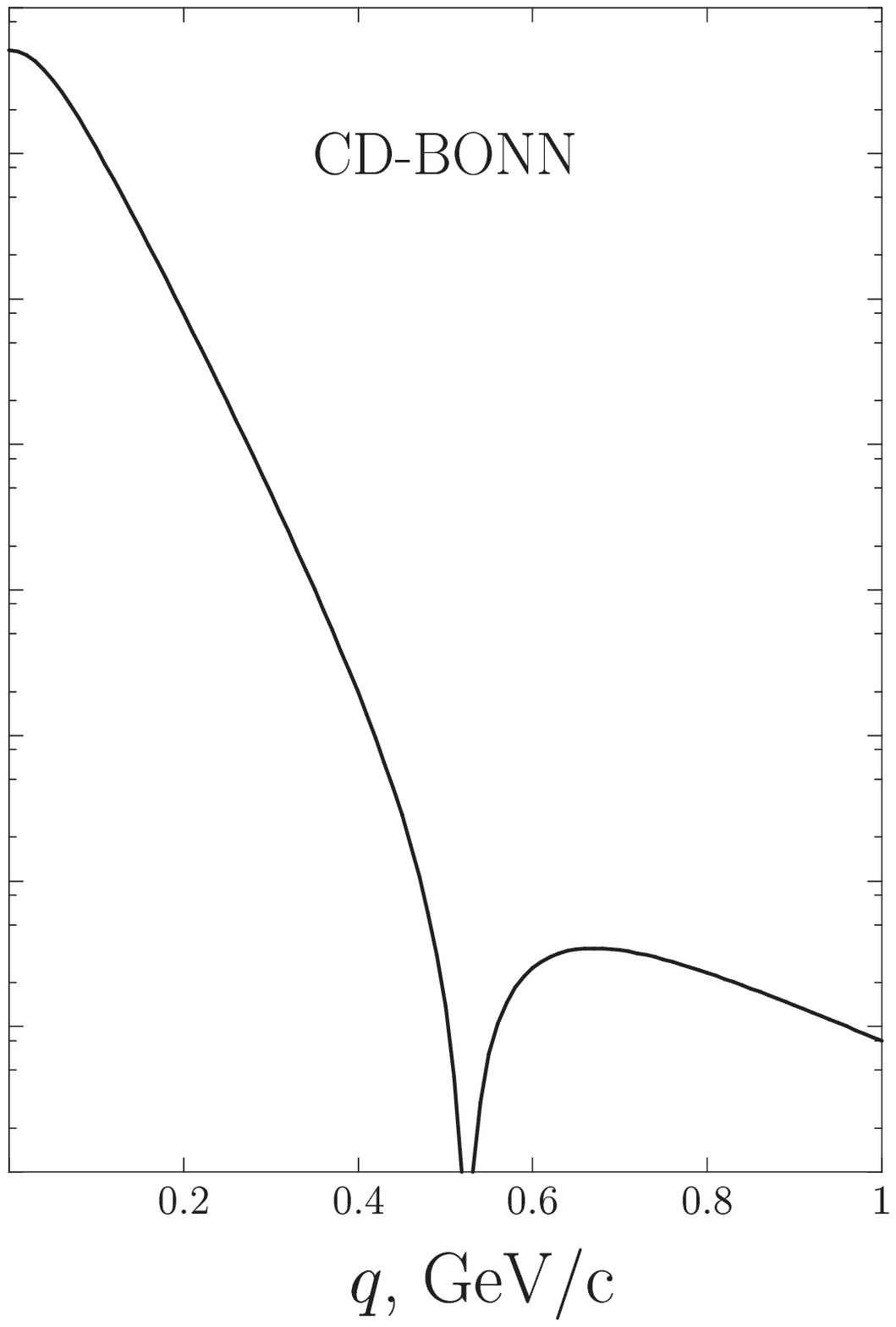}

\vspace{10.cm}
\caption{Momentum distribution of the singlet pair.}
\label{fig:4:sing}
\end{figure*}

In Figure~\ref{fig:2:S+D} the momentum distributions of the triplet pair in
$S$ and $D$ waves are compared with that for the deuteron in $^3\mathrm{He}$. 
The interference momentum distribution and the momentum distribution of the singlet pair
are displayed in Figure~\ref{fig:3:int}~and~\ref{fig:4:sing}, respectively.  

From Figures~\ref{fig:2:S+D}~and~\ref{fig:3:int} one learns that at internal momentum
$q<300$~MeV/c the triplet two-body pair exists mainly as real deuteron, but at higher
$q$ the contribution of nonbound $(np)$ pair becomes essential.

\begin{figure*}
\includegraphics{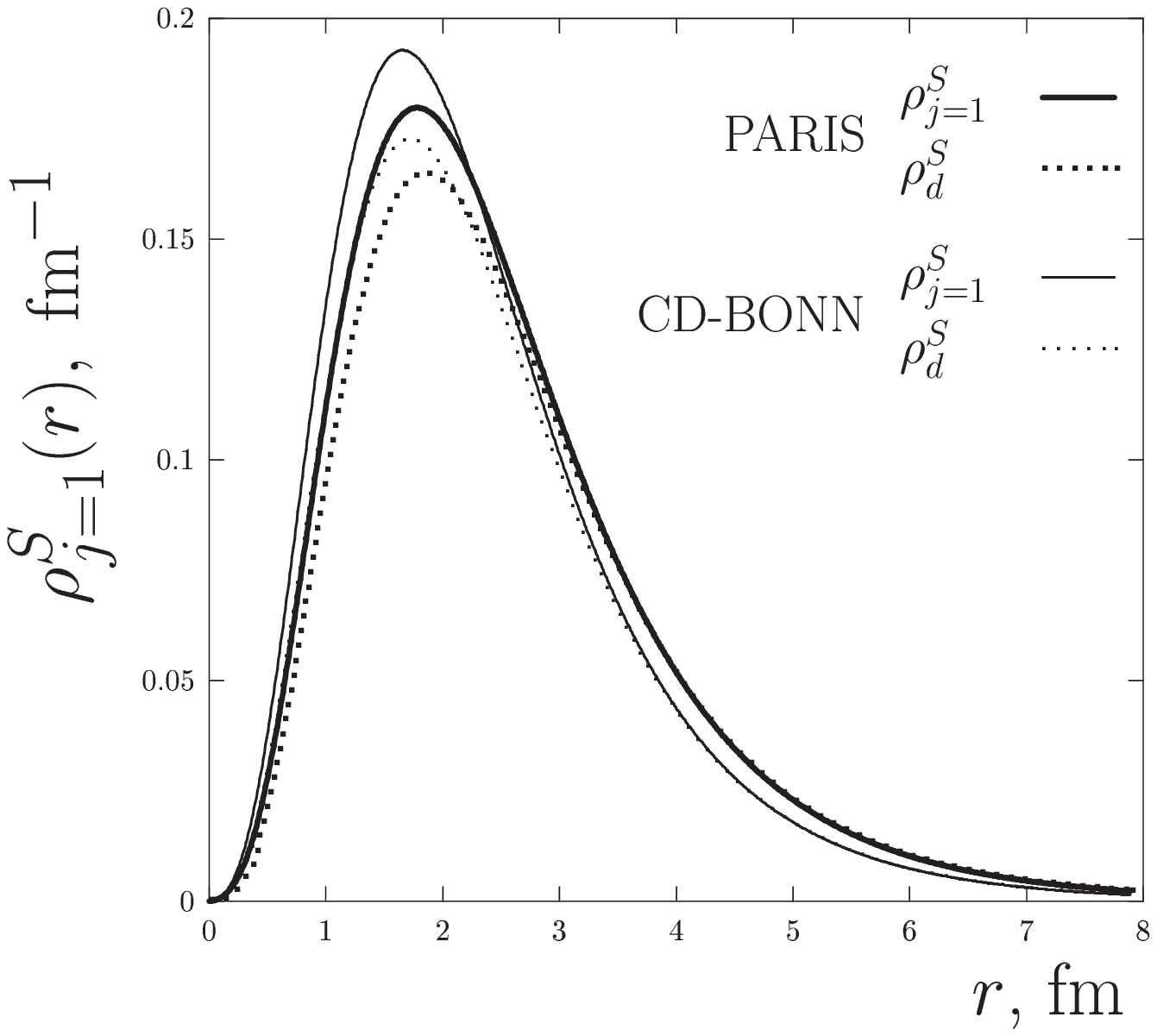}
\includegraphics{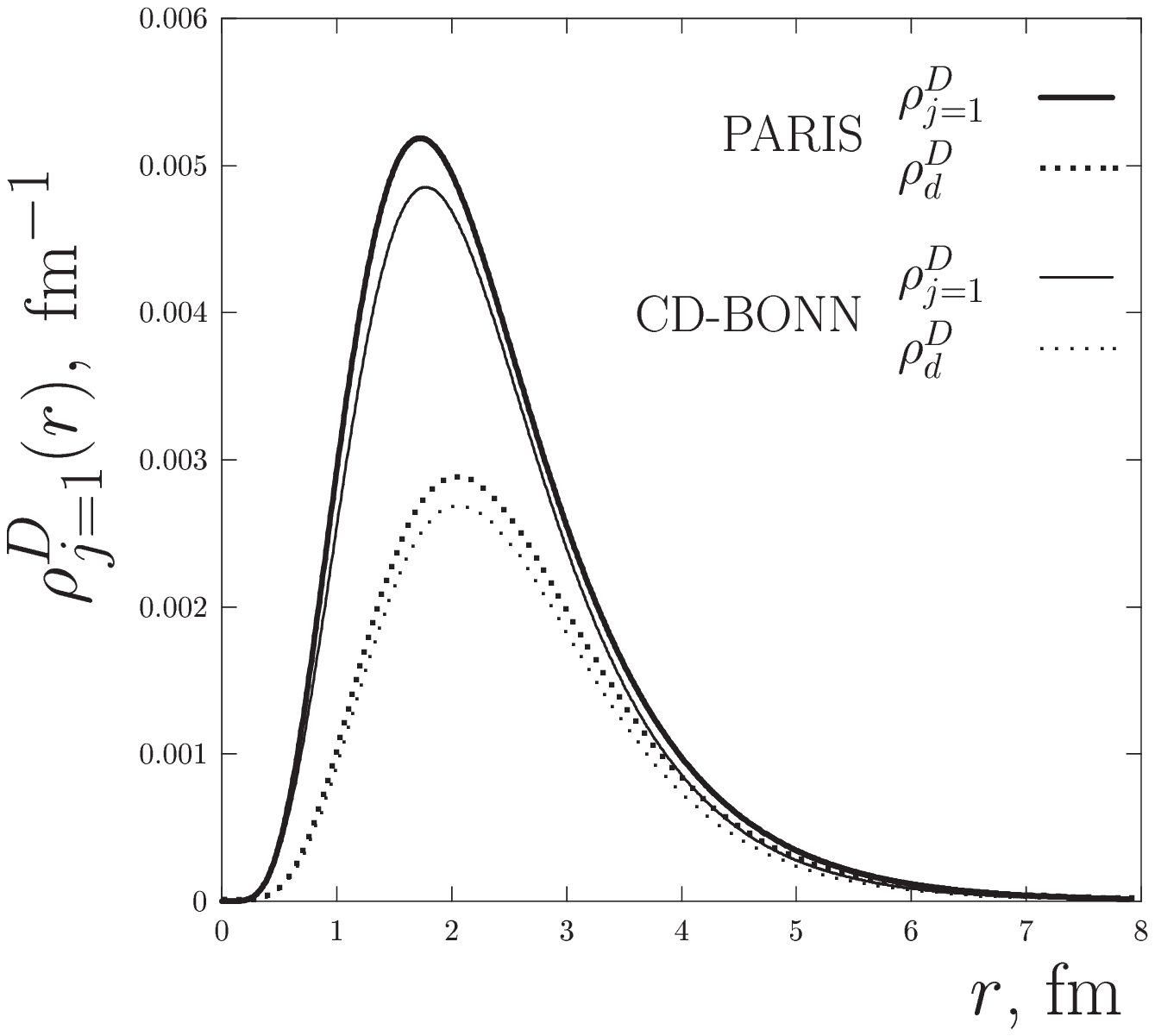}

\vspace{7.cm}
\caption{$S$ (left panel) and $D$ (right panel) density for the triplet $np$ pair.}
\label{fig:4:S+D}
\end{figure*}

\begin{figure*}
\includegraphics{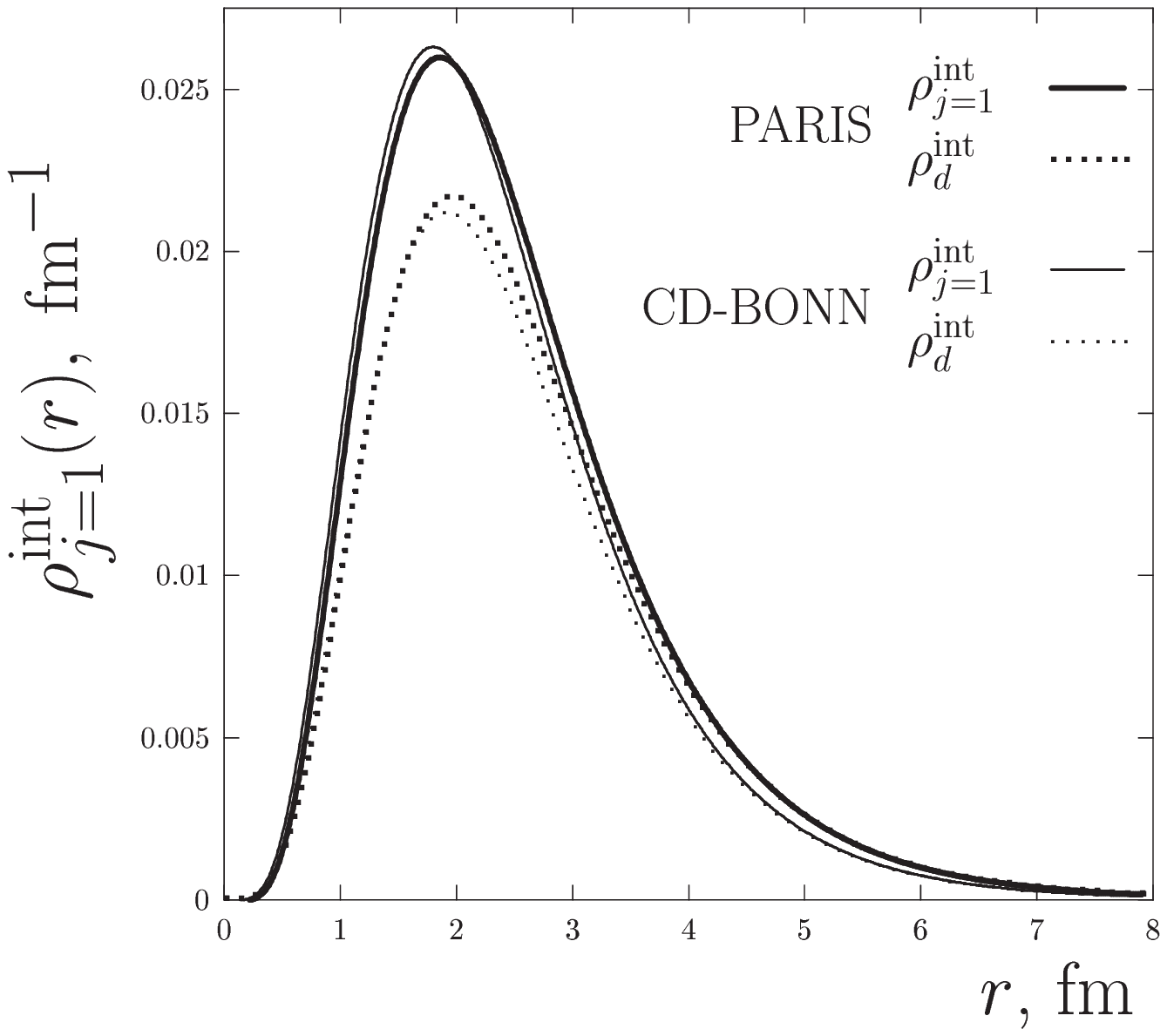}
\includegraphics{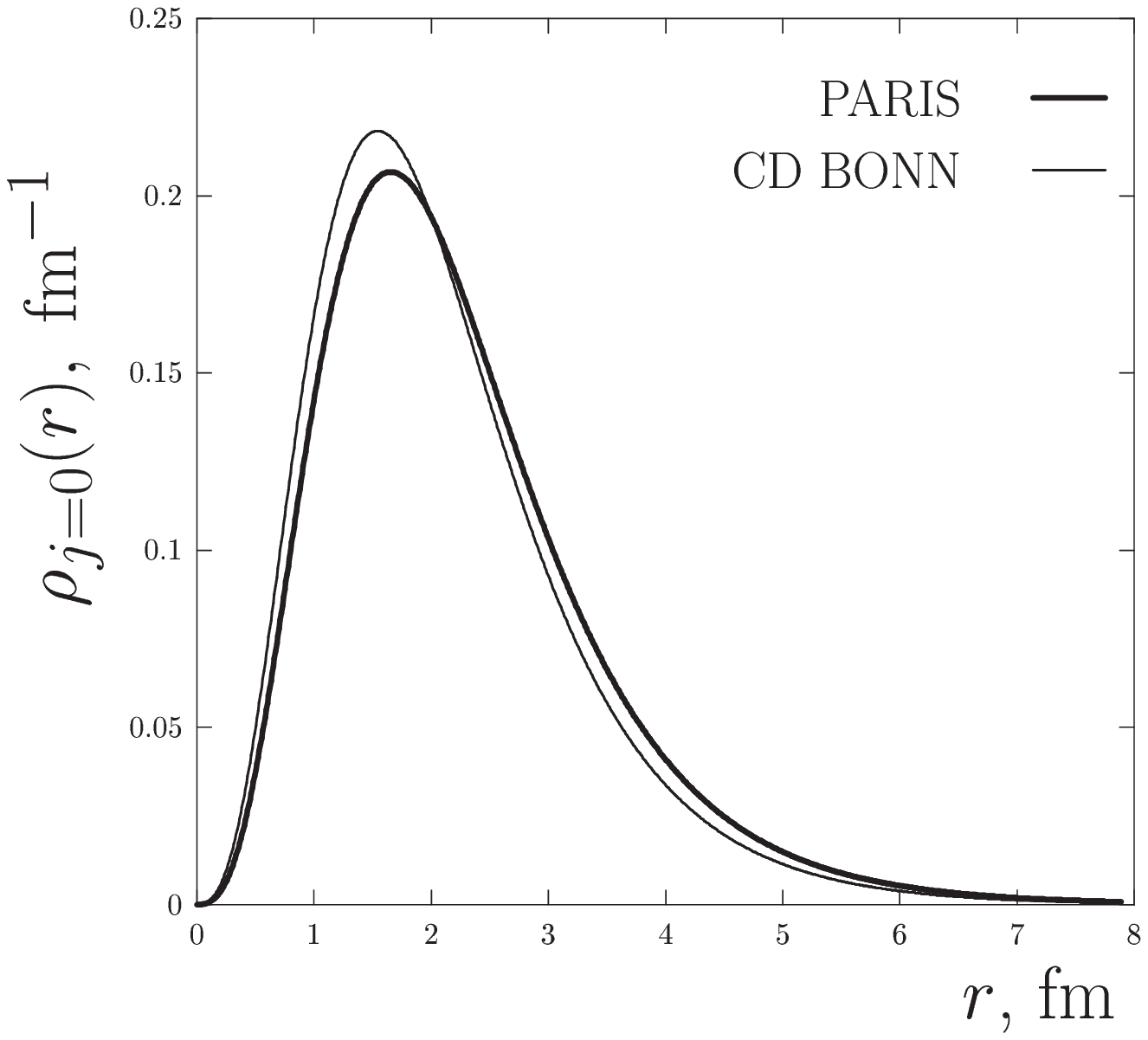}

\vspace{7.cm}
\caption{Interference density for the triplet $np$ pair (left panel) and density for the singlet $np$ pair (right panel).}
\label{fig:5:int+U0}
\end{figure*}

Appropriate two-body densities in configuration space
\bqq
\rho_{3}^S(\rho)&=&3\int_0^\infty\!\!\! \!\! \!dr \left[
\left|\Psi_{^3s_1S}(r,\rho)\right|^2+
\left|\Psi_{^3d_1S}(r,\rho)\right|^2
\right],
\nonumber \\
\rho_{3}^D(\rho)&=&3\int_0^\infty\!\!\! \!\!\!dr \left[
\left|\Psi_{^3s_1D}(r,\rho)\right|^2+
\left|\Psi_{^3d_1D}(r,\rho)\right|^2
\right],
\nonumber \\
\rho_{3}^\mathrm{int}(\rho)&=&3\int_0^\infty\!\!\! \!\! \!dr \left[
\Psi_{^3s_1S}(r,\rho)\Psi_{^3s_1D}(r,\rho)+\right.
\nonumber \\
&&\left.
\ \ \ \ \ \ \
+
\Psi_{^3d_1S}(r,\rho)\Psi_{^3d_1D}(r,\rho)
\right],
\nonumber \\
\rho_{1}(\rho)&=&3\int_0^\infty\!\!\! \!\! \!dr
\left|\Psi_{^1s_0S}(r,\rho)
\right|^2
\label{rho_0}
\eqq
are shown in Figures~\ref{fig:4:S+D} and \ref{fig:5:int+U0}. The triplet densities are also compared
with densities for the deuteron in $^3\mathrm{He}$.


\section{Derivation of reaction amplitude\label{sec:III}}
\subsection{2N-exchange\label{susec:III.B}}
In the nonrelativistic limit and at  $\theta_{\rm cm}=180^{\circ}$
the corresponding amplitudes $A^{\rm (2N)},F^{\rm (2N)}$ and
$G^{\rm (2N)}$ of (\ref{1})  read (see, e.g., \cite{Kob93})
\bqq
A^{\rm (2N)}&=&\frac13 t_3
\left[
n_{3}^S(k) - 2\sqrt2 n_{3}^\mathrm{int}(k)
 + 2 n_{3}^D(k)
\right]
\nonumber \\
&&
\ \ \ \ \  + \frac13 t_1 n_{1}(k),
\nonumber \\
F^{\rm (2N)}&=&\frac13 t_3
\left[
2n_{3}^S(k) + 2\sqrt2 n_{3}^\mathrm{int}(k) +  n_{3}^D(k)
\right],
\nonumber \\
G^{\rm (2N)}&=&\frac13 t_3
\left[
-n_{3}^S(k) + 2\sqrt2 n_{3}^\mathrm{int}(k) - 2 n_{3}^D(k)
\right]
\nonumber \\
&&
\ \ \ \ \  + \frac13 t_1 n_{1}(k),
\label{9.G} 
\eqq
where
$t_j = 8(2\pi)^3 m_\tau m_p \left(
\varepsilon_j -\frac{\vec k^{\,2}}{2 \mu_j}
\right)$ and $\varepsilon_1=m_\tau-2m_p-m_n$, $\varepsilon_3=m_\tau-m_d-m_p$,
$\mu_1=\frac{(m_p + m_n)m_p}{2m_p + m_n}$ and $\mu_3=\frac{m_p m_d}{m_p + m_d}$.

In relativistic case mass of the 2N system becomes indefinite. But we have
estimated mean squared momentum in the pair $\left<p^2\right>$ and find that
at $q<$0.7~GeV/c it is $\left<p^2\right> <0.1\ (\mathrm{GeV/c})^2$. So the effective
mass of the  pair should be close to the deuteron mass. In the forthcoming calculations 
we will use such approximation. 

Another kind of relativistic effects comes from relativistic deformation of the
internal dynamics in the bound state considered in the infinite momentum 
frame (IMF). For elastic scattering with rearrangement of clusters,
application of the IMF dynamics needs a care, because reaction amplitude can 
lose symmetry under initial and final states \cite{Kondratyuk81}. 
This problem was also considered in \cite{Kob86} and the appropriate formalism
was proposed. Assuming relativistic invariance of the expression
\beq
\Psi=\frac\Gamma{p_\mathrm{cl}^2 + m_\mathrm{cl}^2}
\label{IMF:Psi_Gamm}
\end{equation} 
where $\Psi$ is the wave function, $\Gamma$ is the vertex function for
$^3\mathrm{He}\to (np)+p$ virtual decay amplitude, $p_\mathrm{cl}$ and
$m_\mathrm{cl}$ are the cluster momentum and mass, one can consider the initial and
final $^3\mathrm{He}$ in their \underline{``own IMF''}. This infinite momentum frames 
are defined to be limiting frames in which observer is moved with velocity close
to the speed of light in the direction opposite to the motion of the every 
$^3\mathrm{He}$ in the reaction center of mass frame. As a result, the 
invariance
under initial and final states is restored: the wave function for the 
initial, as well as for the final, $^3\mathrm{He}$ will 
depend \underline{on the same} light cone variable
\beq
\alpha=\frac{E_p^{\ast} + p^{\ast}}{E_\tau^{\ast} + p^{\ast}},
\label{IMF:Psi}
\end{equation}
where the energies $E_p^{\ast}$, $E_\tau^{\ast}$ and the momentum $p^{\ast}=|\vec p^{\,\ast}|$ are defined in the reaction center of mass frame.

After that the ``relativization procedure'' is reduced to the two prescriptions:
\begin{itemize}
\item substitute new argument in (\ref{9.G}) 
$
k \to k_{\rm IMF}
$
\item change the factor
$
t_1\approx t_3 \to 4(2\pi)^3
\frac{\epsilon_p \epsilon_d
\left(
M^2_{dp}-m^2_\tau
\right)}
{(\epsilon_p + \epsilon_d )(1-\alpha)}
$
\end{itemize}
In these prescription the relativistic internal momentum, $k_{\rm IMF}$, the
invariant mass of virtual $d+p$ pair, $M_{dp}$, and other variables are expressed as follows:
\bqq
k_{\rm IMF}&=&\sqrt{\frac{
\lambda (M^2_{dp},m^2_d,m^2_p)
}
{4 M^2_{dp}}},
\label{12.1} \\
M^2_{dp}&=&\frac{m^2_p}\alpha + \frac{m^2_d}{1-\alpha},\\
\epsilon_p &=& \sqrt{m^2_p + k^2_{\rm IMF}},\
\epsilon_d = \sqrt{m^2_d + k^2_{\rm IMF}},
\eqq
where $\alpha$, defined in (\ref{IMF:Psi}), is
the fraction of the $^3\rm He$ momentum carried by the 
proton in IMF.

\subsection{High momentum transfer by intermediate pion\label{sec:III.B}}
The matrix element corresponding to PI-diagram of Figure~\ref{fig:1} reads:
\begin{widetext}
\bqq
{\mathcal M}_{mM}^{m'M'}({\rm PI}) =-
3\left(\frac1{2\pi}\right)^8
\left(2m_p\frac{f_{\pi NN}}{\mu}\right)^2
\int d^4p_d d^4p_d'\sum_{\sigma \sigma '}
A_{\sigma \sigma '} F_\pi^2(q^2) 
 \overline u_{m'}(p') \gamma_5
\frac1{{l \!\! /}' -m_p +i0}\Gamma_\mu
\varepsilon^\mu(\sigma')U_M(P)
\nonumber \\
\times 
\overline U_{M'}(P')\Gamma^\nu \varepsilon^{\ast}_\nu(\sigma)
\frac1{{l \!\! /} -m_p +i0}\gamma_5u_m(p)
 \frac1{(p_d^2 - m_d^2+i0)(p_d{'}^2 - m_d^2+i0)
(q^2-\mu^2 +i0)(q{'}^2 - \mu^2 +i0)}, 
\label{PI.1}
\eqq
where $A_{\sigma \sigma'}$ is an amplitude of the subprocess $\pi^0 d \to \pi^0 d$,
$\mu$ is the pion mass,
$F_\pi(q^2)$ and $f_{\pi NN}$ are the form factor and coupling constant of the $\pi NN$
vertex. 
In (\ref{PI.1}) $\Gamma_{\mu}$ is the virtual $^3{\rm He}\to d+p$ decay amplitude
\beq
\frac{\overline u_m(p) \Gamma_{\mu}\varepsilon^\mu (\sigma) U_M(P)}
{p^2-m_p^2 +i0}
=(2\pi)^\frac32 \sqrt {2 m_d}\,\, \psi_M^{\sigma m}(\vec k).
\label{5}
\end{equation}
Here and later on we use the following notations: $p$,  $P$ are momenta and $m$, $\sigma$,
$M$ are magnetic quantum numbers of the proton, the deuteron and
$^3{\rm He}$, respectively;
$\vec k= \frac23 \vec P - \vec p_d$ is the relative momentum
between the proton and the deuteron in $^3{\rm He}$;
$\varepsilon_3$ and $\mu_3$ were already defined in Sec.~\ref{susec:III.B};
$\varepsilon^\mu (\sigma)$ is the polarization vector of the deuteron,
$u_m(p)$ and $U_M(P)$ are spinors for the proton and $^3{\rm He}$;
$\psi_M^{\sigma m}(\vec k)$ is the overlap between the $^3{\rm He}$ and $p+d$
wave functions. The spinors are normalized as $\overline u_m(p) u_{m'}(p)=2m_p\delta_{mm'}$, etc. 

Integrating over the deuteron energies one gets
\bqq
{\mathcal M}_{mM}^{m'M'}({\rm PI}) =
3\left(\frac1{2\pi}\right)^6
\left(2m_p\frac{f_{\pi NN}}{\mu}\right)^2 2m_d (2\pi)^3
\int \frac{d^3p_d}{2E_d}\frac{d^3p_d'}{2E_d'}\sum_{\sigma \sigma '}
A_{\sigma \sigma '} 
F_\pi^2(q^2) \chi^\dag_{m'}\vec Q' \vec \sigma \chi_{\tilde m'}
 \chi^\dag_{\tilde m}\vec Q \vec \sigma \chi_{m}
\nonumber \\
\times
\frac{
\psi_{M'}^{\ast\sigma \tilde m}(\vec k{'})
\psi_{M}^{\sigma '\tilde m'}(\vec k)
}{
(q^2-\mu^2 +i0)(q{'}^2 - \mu^2 +i0)
},
\label{PI.14}
\eqq
\end{widetext}
where $\chi_{m(m')}$ and $\chi_{\tilde m(\tilde m')}$ are Pauli spinors for the protons,
$\vec Q= \sqrt{\frac{E_{p} + m_p}{E_{l} + m_p}}\vec l -
\sqrt{\frac{E_{l} + m_p}{E_{p} + m_p}}\vec p$,
$\vec Q'= \sqrt{\frac{E_{p'} + m_p}{E_{l'} + m_p}}\vec l' -
\sqrt{\frac{E_{l'} + m_p}{E_{p'} + m_p}}\vec p{\,'}$
 and
$E_d=\sqrt{\vec p_d^{\,2} + m_d^2}$, $E_d'=\sqrt{\vec p_d^{\,\prime 2} + m_d^2}$.

To simplify the loop integration we take out of the integral the amplitude $A_{\sigma \sigma'}$ and the
form factors $ F_\pi^2(q^2)$ at point where the deuteron carries $\frac23$ of the $^3{\rm He}$
momentum.  To take into account Fermi motion of the deuteron in $^3{\rm He}$,
these factors were averaged over Gaussian distribution with
$\left<p_d^2\right>^{1/2}=$41.3~MeV/c. The latter
value was taken from the calculated deuteron momentum distribution 
in $^3{\rm He}$.
Finally one gets:
\begin{widetext}
\bqq
{\mathcal M}_{mM}^{m'M'}({\rm PI})& =
&\frac{6}{m_d(2\pi)^3}
\left(2m_p\frac{f_{\pi NN}}{\mu}\right)^2  \sum_{\sigma \sigma '}
\left<A_{\sigma \sigma '} F_\pi^2\right> 
\chi^\dag_{m'} \sigma_j \chi_{\tilde m'}
 \chi^\dag_{\tilde m} \sigma_i \chi_{m}
{\mathcal Q}_{jM'}^{\sigma \tilde m}
{\mathcal Q}_{iM}^{\sigma' \tilde m'},
\label{PI.20} \\
{\mathcal Q}_{jM'}^{\sigma \tilde m}&=&
\int \frac{d^3p_d Q_j\psi_{M'}^{\ast\sigma \tilde m}
(\vec k)}{2(q^2-\mu^2 +i0)},\ \
{\mathcal Q}_{iM}^{\sigma' \tilde m'} =
\int \frac{d^3p'_d Q'_i
\psi_{M}^{\sigma' \tilde m'}(\vec k{'})}{2(q{'}^2-\mu^2 +i0)},
\label{PI.20.b} 
\\
\left<A_{\sigma \sigma '} F_\pi^2\right> &=& \frac{1}{\sqrt{\pi\left<p_d^2\right>}}
\int_{-\infty}^{\infty} d\tilde p e^{-\frac{(\tilde p - p^\ast)^2}{\left<p_d^2\right>}}
A_{\sigma \sigma '}(\tilde s_{\pi d}, 180^\circ) F_\pi^2(\tilde q^2).
\label{PI.20.c}
\eqq
\end{widetext}
Now there are two independent three-dimensional integrals and in the nonrelativistic limit the
integration over angles can be done analytically.

\subsection{Direct mechanism in optimal approximation}
For DIR mechanism we use optimal approximation which minimizes
the binding energy and recoil corrections \cite{Gurvitz_et_al}:
\bqq
\mathcal{M}^{m{'}M{'}}_{mM}({\rm DIR}) 
&=& 3 \sum_{\sigma{,}m_1{,}m_1{'}}
M^{m{'}m_1{'}}_{mm_1} 
\nonumber \\
&&\!\!\!\!\!\!\!\!\!\!\!\!\!\!\!\!\!\!\!\!\!\!\!\!\!\!\!\!\!\!\!\!\!\!\!\!\!\!\!\!
\times
\sum_{\nu \nu'}
\int d^3pd^3q \Psi^{\nu'\,*}_{M{'}\sigma m_1{'}}(\vec p,\vec q\,')
\Psi^\nu_{M\sigma m_1}(\vec p,\vec q).
\label{DIR.1}
\eqq
The meaning of the quantum numbers $M,\sigma,m$ etc. is clear from the DIR diagram
(Figure~\ref{fig:1}).
The amplitude $M^{m{'}m_1{'}}_{mm_1}$ is the on-shell $pp$ elastic
scattering amplitude at effective energy,
$M^{m{'}m_1{'}}_{mm_1}(E_\mathrm{eff},\theta)$, where $E_\mathrm{eff}$
corresponds to such total
energy in the Breit frame as if the struck proton takes all the
momentum of  $^3\rm He$ \cite{Gurvitz_et_al}.

At $\theta_{cm} = 180^{\circ}$ the $ pp$-amplitude has three
independent spin amplitudes~\cite{Lehar}:
\beq
M^{m{'}m_1{'}}_{mm_1} =
\left(
\begin{array}{c c c c}
(a+d) & 0 & 0 & 0\\
0 & (a-c) & (a-b) & 0\\
0 & (a-b) & (a-c) & 0\\
0 & 0 & 0 & (a+d)
\end{array}
\right){,}
\label{DIR.2}
\end{equation}
with the constraint $a(\pi) - b(\pi) = c(\pi) -d(\pi)$.

Thus in the optimal approximation DIR amplitudes read:
\begin{eqnarray}
A^{\rm DIR} &=& \left(a + \frac{d-2c}{3}\right) F_0(p) + \frac{c+d}{3}
F_2(p), \nonumber \\
F^{\rm DIR} &=& \left(a + \frac{2d - c}{3}\right) F_0(p) - \frac{c+d}{3}
F_2(p), \\
G^{\rm DIR} &=& -\frac{a - b}{3}\left[F_0(p) + F_2(p)\right], \nonumber
\label{DIR.3}
\end{eqnarray}
where
\begin{eqnarray}
F_0(p) &=&3 \int_0^{\infty} dr j_0(pr)\left[u^2(r) +w^2(r)\right],
\nonumber \\
F_2(p) &=&3 \int_0^{\infty} dr j_2(pr)\left[w^2(r) + 2\sqrt{2}u(r)w(r)\right]
\label{DIR.4}
\end{eqnarray}
and $p=\frac{2}{3}q$, $q^2 = -(p-p{'})^2$.

Note that at high energy $a \simeq c$ and $d \simeq b \simeq 0$ so that
\begin{eqnarray}
A^{\rm DIR} &=& \frac{a}{3}\left[ F_0(p) + F_2(p)\right], \nonumber \\
F^{\rm DIR} &=& \frac{a}{3}\left[ 2F_0(p) - F_2(p)\right]. \\
G^{\rm DIR} &=& -A^{\rm DIR} \nonumber
\label{DIR.5}
\end{eqnarray}

For the $pp$ elastic scattering amplitudes $a$, $b$, $c$ and $d$ at
$E_\mathrm{eff} \leq 1300$~MeV we
have used results of the partial wave analysis by Saclay-Geneva
group~\cite{Lehar}. At higher energy the "diffractive"
parametrization was used:
$
a = c = \frac{p^\ast}{4\pi} (i + \rho_{pp}) \sigma^{\rm tot}_{pp}
$, $d = b = 0$,
where $\rho_{pp}$ is the ratio of the real to imaginary part of the forward
scattering amplitude and  $\sigma^{\rm tot}_{pp}$ is the total cross
section of the $ pp$-scattering.

\section{Numerical calculations, comparison with experiment and predictions
\label{Numeric}}
The standard parametrization of the form factor
$F_\pi (q^2)=(\Lambda^2-\mu^2)/(\Lambda^2 - q^2)$ with
$\Lambda=1300$~MeV and $f^2_{\pi NN}/4\pi = 0.08$ \cite{Machleidt}
was used. We have also provided calculations with another cutoff parameter $\Lambda$,
varying it from 600 to 1700 MeV, but the results were not changed significantly.

Amplitudes $A_{\sigma \sigma'}$ are taken from the partial wave
analysis by Virginia group \cite{Arndt}. It must be emphasized here, that
in most of the previous studies (see
\cite{Uzikov98,CWilkin,KolybSm,Barry,KondrLev,Nakamura}) people either built
special theoretical models for the subprocess or made simplifying approximations
to replace the amplitude by experimental data on the corresponding cross sections.
It is proved by experience that such procedures are not satisfactory.

\begin{figure*}
\includegraphics{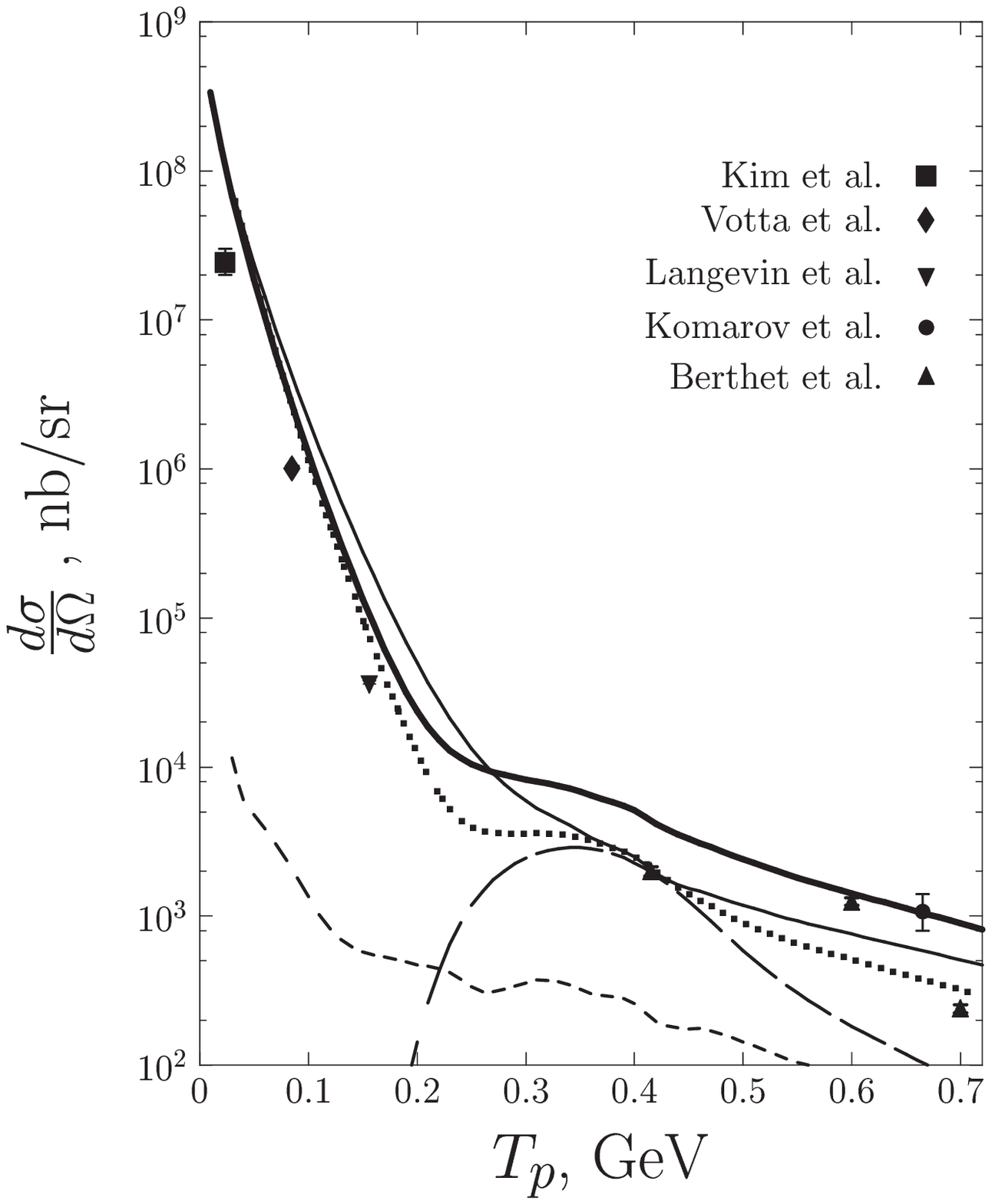}
\includegraphics{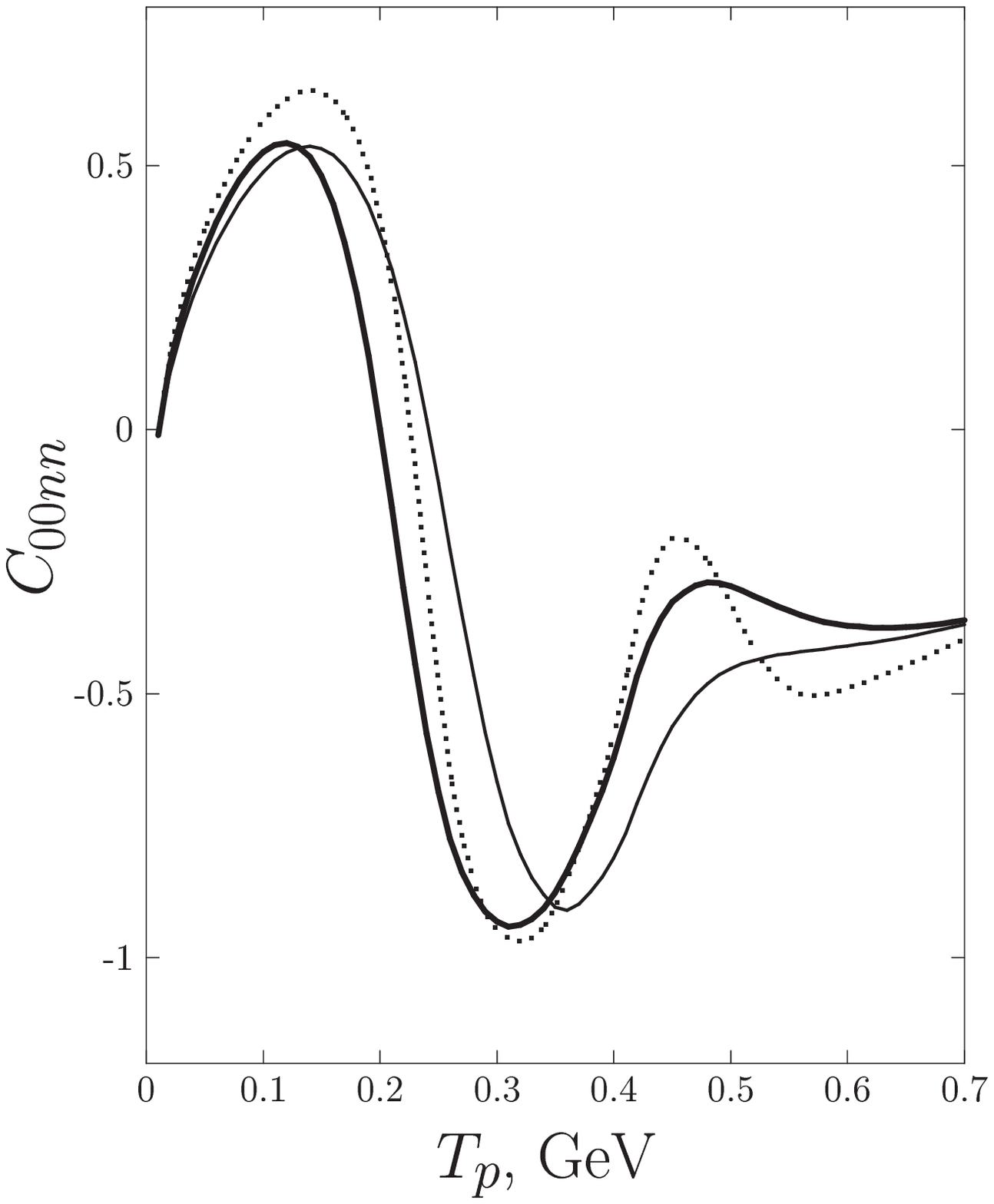}

\vspace{12.cm}
\caption{The differential cross section (left panel) and polarization correlation $C_{00nn}$ (right panel) of the elastic proton-$^3\mathrm{He}$ scattering at $\theta_\mathrm{c.m.}=180^\circ$. The solid curves include $(np)_{1}+(np)_{3}$ exchange together with PI-mechanism (the bold and thin curves are for Paris and CD-Bonn potentials, respectively). The one-deuteron-exchange together with DIR and PI mechanisms is shown by dotted line, the contribution of PI and DIR mechanismin the diferential cross section  is shown by the long- and short-dashed line, respectively; both are for Paris potential. Data are from \cite{Kim}-\cite{SATURNE}. Data \cite{Kim}-\cite{Langev} were extrapolated to $\theta_\mathrm{c.m.}=180^\circ$ by us. The point \cite{Komarov} is at $\theta_\mathrm{c.m.}=169^\circ$.}
\label{fig:cs+Cyy}
\end{figure*}

Results of the calculations at $T_p<$~700~MeV are shown in Figure~\ref{fig:cs+Cyy}. 
At the left panel we compare our results with  experimental data for
the differential cross section. Predictions for the polarization correlation 
$C_{00nn}$ are given in the right panel of Figure~\ref{fig:cs+Cyy}. DIR 
mechanism gives sizable contribution  only at high energy ($T_p>1$~GeV). 

\section{Conclusions and remarks \label{Conclusions}}
The main results of this work can be summarized as follows:
\begin{itemize}
\item The theoretical description of the proton-$^3{\rm He}$ EBS at $T_p<700$~MeV
is given. The energy dependence of the differential cross section is described and 
the polarization correlation $C_{00nn}$ is calculated. The model predicts a 
structure in energy dependence of the differential cross section and 
spin-dependent observables between 200 and 400~MeV, which comes from the 
interference between 2N-exchange and PI-mechanism. Its verification in 
experiment is now possible and important for understanding mechanism of 
EBS on the lightest nuclei. Such experimental program is now in preparation in RCNP.
\item Only at low energy ($T_p\lesssim$~150~MeV) a cluster two-body approximation
$d+p$ for $^3\mathrm{He}$ can be justified (see difference between full and dotted
curves in Figure~\ref{fig:cs+Cyy}). At higher energy total three-body structure of
$^3{\rm He}$, including scattering state in triplet and singlet 2N pair, becomes of great
importance.
\item At $T_p\gtrsim$~700~MeV additional mechanisms should be taken into account.
Between them one can mention DIR-mechanism and sequential transfer of 
noninteracting $np$ pair \cite{BLU}. The later is defined by high relative
momentum in the pair, $p>$~600~MeV/c, and low spectator momentum $q\sim$~100~MeV/c.
Such high internal momentum in the pair corresponds to a picture of overlapping 
nucleons and $^3\mathrm{He}$  should be considered rather as a 9-quark system, 
than a trinucleon bound state.
\end{itemize}

We do not consider distortion effects in this paper. Such effects were estimated 
on the basis of Glauber-Sitenko theory \cite{Uzikov98}, but application
of such approach at the RCNP energy (as it was done in the recent paper \cite{Uzikov02})
is very doubtful. This problem, as well as angular dependence of the 
reaction observables, will be considered in our forthcoming publication.

\begin{acknowledgments}
The authors would like to thank M.~Tanifuji and H.~Toki for stimulating
discussions. One of us (A.P.K.) is grateful to A.~Boudard for sending
the numerical cross section data obtained at SATURNE. We also indebted
to F.~Lehar for valuable discussion of the partial wave analysis 
of the $pp$ elastic amplitude  by Saclay-Geneva group and for the numerical
tables of the partial waves. The authors are thankfull to I.I.~Strakovsky for a lot of useful
discussions of the results of partial wave analysis by Virginia group. Two 
of the authors  (A.P.K., E.A.S.) acknowledge the hospitality of RCNP,
where this work was carried out with a Center of Excellence grant from the
Ministry of Education, Culture, Sports, Science and Technology
(Monbu-Kagaku-sho), Japan.
\end{acknowledgments}

\end{document}